\documentclass{ceab}   

\usepackage{epsfig}     
\usepackage{graphicx}   

\usepackage{ceabbib}     
\usepackage[T1]{fontenc}

\def\gore{HXR coronal sources during decay phase of solar flares}

\begin{document}

\title{RHESSI Investigation of X-Ray Coronal Sources During Decay Phase of Solar Flares: I. Observations}
%
\author{T.Mrozek, Z. Ko\l tun, S.Ko\l oma\'nski and U.B\c ak-St\c e\' slicka        
\vspace{2mm}\\
\it Astronomical Institute, Wroc{\l}aw University, Poland }

\maketitle

\begin{abstract}
We analyse the observational characteristics for a set of Long Duration Events (LDE) using Reuven Ramaty High Energy Solar Spectroscopic Imager ({\it RHESSI}). Excellent energy resolution (1 keV) of HXR images reconstructed with {\it RHESSI} allowed us to perform imaging spectroscopy of coronal loop top sources (LTS) observed for hours after maximum of the flare. We found that the sources are large structures. The diameter of an LTS is changing with time and energy. High temperature component can be observed for few dozen hours after the maximum. For some time intervals clear non-thermal component was observed in spectra, but there was no significant footpoint emission at that time. All thermal and non-thermal emissions originate from the LTS. The obtained observational characteristics were used in calculating the energy balance within LTS, which are presented in \citet[]{kolomanski2011}(this issue).

\end{abstract}


\section{Introduction}

\citet{veselovsky2008} analysed duration times for over 40000 flares observed by {\it GOES} satellites for years 1976 - 2006. The authors found that the histogram of durations can be easily described with the use of lognormal distribution. With such a distribution it is possible to divide flares into three groups: impulsive - shortest duration, typical - located close to the maximum of the histogram and long duration events (LDE) - located to the right of the maximum. Such a division is arbitrary. There is no difference in the physical mechanisms responsible for each group as these flares belong to one distribution with no significant fluctuations. As at the basis there is no difference between physics of solar flares from the three groups then the investigation of LDEs gives better insight into the nature of the energy release process with present instrumentation. 

The decrease of brightness of LDEs observed in soft X-rays (SXR) may last from several hours to more than a day. Much insight into the nature of LDEs was made by ultraviolet and X-ray  observations, during the \textit{Skylab}, \textit{SMM} (Solar Maximum Mission) and \textit{Yohkoh} space missions \citep[e.g.][]{sheeley1975, kahler1977, feldman1995, murnion1998, czaykowska1999, shibasaki2002, isobe2002}. One of the most important conclusions is that without the continuous energy input during the whole decay phase the LDEs would decay much faster than it is observed.

The whole X-ray emission of LDEs is located in loop-top sources (LTSs) which form before the flare maximum and last for many hours \citep[e.g.][]{feldman1995, kolomanski2007a}. LTSs are very promising features for an analysis of the energy release during the decay phase as they should be close to the primary energy release site \citep[e.g.][]{kopp1976, shibata1999, hirose2001, karlicky2006}. Moreover, the presence of LTSs during the whole flare decay-phase requires continuous energy release and some restriction mechanism efficiently preventing outflow of mass and energy from them \citep[see][]{vorpahl1977}. Observations show that without such mechanism LTSs would rapidly lose energy by radiative and conductive processes and should vanish within minutes \citep[e.g.][]{jiang2006, kolomanski2007b}.

Hard X-ray LTSs were rarely observed for hours after flare maximum. \citet[]{masuda1998} observed HXR sources up to $30$ minutes after flare maximum. The sources were large ($1-2$~arcmin in diameter). Two other LDEs were investigated by \citet[]{murnion1998}. The sources observed grew with time and had diameters  $\approx20-45$~arcsec. In one of these flares the HXR source was visible up to 3 hours after flare maximum. The analysis of three LDEs with the use of {\it Yohkoh} data was made recently by Ko\l oma\'nski (2007a, b). The author reported that HXR sources ($14-23$~keV) were observed for about 50 minutes after flare maximum. The HXR source existing more than 1 hour after flare maximum was also reported by \citet[]{khan2006}. In all these cases the HXR emission was observed close to and above the SXR emission source. 

First investigation of long persisting LTS based on {\it RHESSI} data was made by \citet[]{gallagher2002}. The HXR emission was observed in the $12-25$~keV range some 4 hours after flare maximum and almost 11 hours in the $6-12$~keV range. The LTS was large and its altitude increased at a speed that gradually declined from $10$ to $1.7\;{\rm km}\:{\rm s}^{-1}$. The higher energy emission ($12-25$~keV) was located above the lower energy emission  ($6-12$~keV and $3-6$~keV) and the whole HXR emission was located above tops of the loops observed in the EUV range ({\it TRACE}~195~\AA). Another long persisting LTS has been reported recently \citep{sainthilaire2009}. The authors observed an LTS in the energy range $6-12$~keV even 12 hours after the flare maximum. The source observed was spatially correlated with the FeXVIII emission observed by Ultra Violet Coronograph Spectrometer (UVCS) onboard {\it SOHO}. 

{\it Yohkoh}/HXT and {\it RHESSI} data have provided useful information about LDEs, but only in one channel significant emission was observed. Poor energy resolution of {\it Yohkoh}/HXT was caused by the instrument characteristics. Despite the fact that {\it RHESSI} has excellent energy resolution, the whole previous analysis was restricted to wide energy intervals which still do not distinguish the actual nature of long persisting LTS. Here we present the investigation made with the use of {\it RHESSI} images reconstructed in very narrow ($1$~keV) energy intervals. {\it RHESSI} is very useful for analyzing weak sources even with 1~keV energy resolution due to its high sensitivity. Thus, it gives an opportunity to determine the actual nature of emission of LTSs (thermal, non-thermal). Using the images we estimate physical parameters of LTSs through imaging spectroscopy. 

\begin{table}[t]
\caption{\footnotesize{\textit{List of analyzed flares. (1) - Date, (2) - {\it GOES} class, (3) - Heliographic coordinates, (4) - {\it GOES} maximum [UT], (5) - {\it GOES} decay time [h] .}}\label{tab1}}
\begin{center}
\begin{tabular}{cccccc}
\hline
\noalign{\smallskip}
 & (1) & (2) & (3) & (4) & (5) \\
\noalign{\smallskip}
\hline
\noalign{\smallskip}
1 & 25/10/2002 & M1.5 & N36W09 & 17:47 & 12 \\
2 & 25/8/2003 & C3.6 & S11E41 & 02:59 & 7 \\
3 & 11/11/2003 & C8.5 & N00E89 & 16:16 & 13 \\
4 & 05/1/2004 & M6.9 & S05E57 & 03:45 & 34 \\
5 & 20/1/2005 & X7.1 & N18W74 & 07:01 & 48 \\
6 & 30/7/2005 & X1.3 & N10E59 & 6:36 & 11 \\
7 & 22/8/2005 & M2.7 & S10W52 & 01:34 & 11 \\
8 & 29/11/2005 & C4.0 & S14W45 & 17:09 & 8.5 \\
9 & 25/1/2007 & C6.3 & S07E90 & 07:15 & 17 \\\hline
\end{tabular}
\end{center}
\end{table}

\begin{figure}[!t]
\begin{center}
\epsfig{file=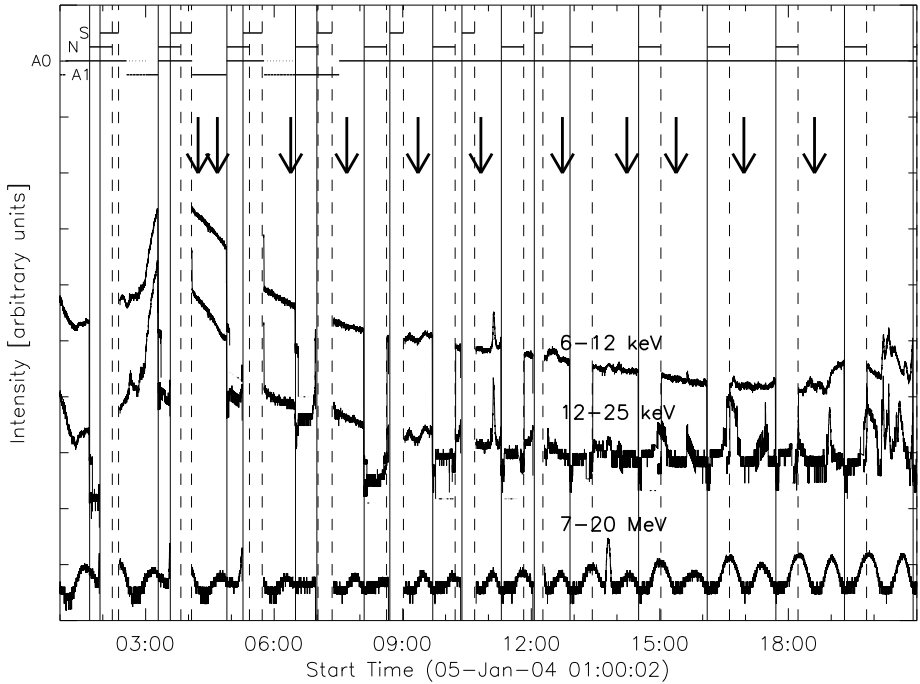,width=12cm}
\end{center}
\caption{{\it RHESSI} light curves of the entire event no.~4. Vertical lines represent borders of the spacecraft night (N) and the SAA (S) periods. Arrows mark time intervals in which the analysis was performed. \label{fig1}}
\end{figure}

\begin{figure}[!t]
\begin{center}
\epsfig{file=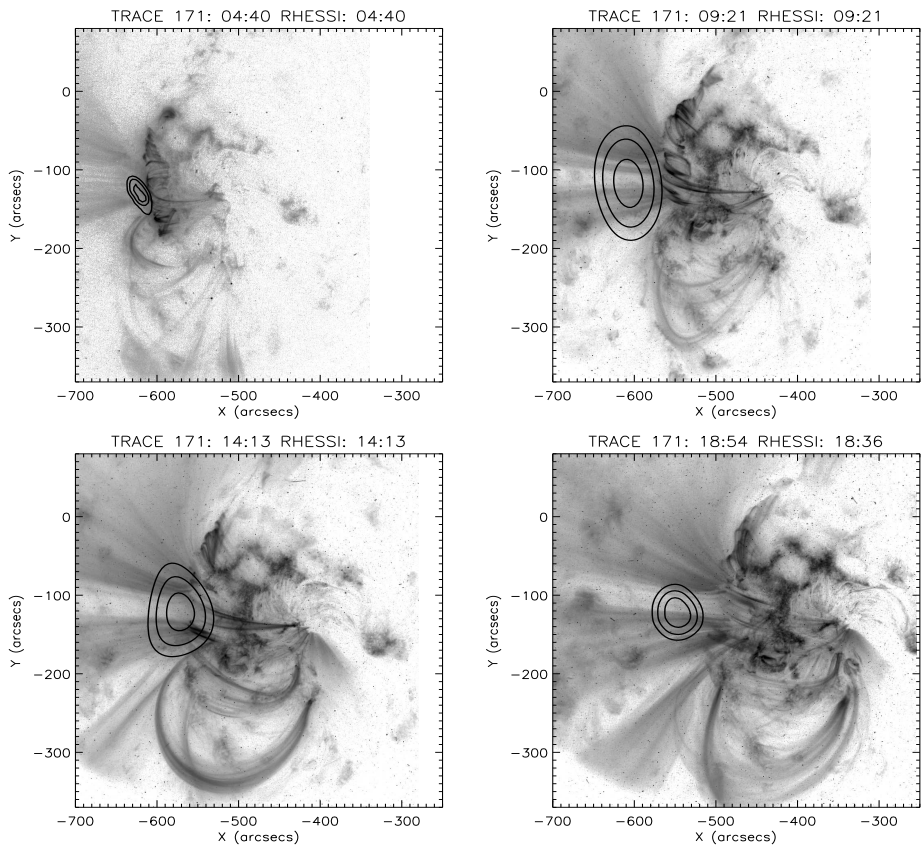,width=12cm,height=11cm}
\end{center}
\caption{{\it TRACE} images (171 \AA) with {\it RHESSI} contours ($6-7$ keV) for the 2004 January 5 flare. \label{fig2}}
\end{figure}

\begin{figure}[!t]
\begin{center}
\epsfig{file=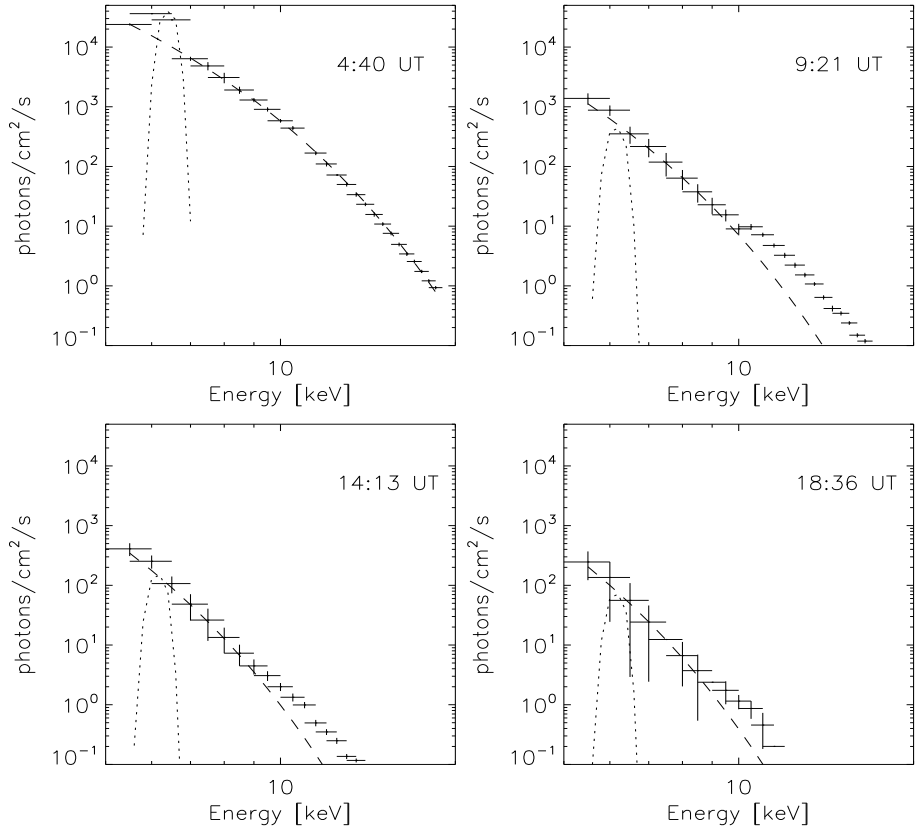,width=12cm,height=11cm}
\end{center}
\caption{Spectra obtained for the 2004 January 5 flare. Times are the same as in Fig.~\ref{fig2}. Thermal component and 6.7 keV line fits are shown. The excess observed for energies greater than 10~keV has been fitted with power-law function. Obtained slopes are high, of the order of 10.\label{fig3}}
\end{figure}

\begin{figure}[!t]
\begin{center}
\epsfig{file=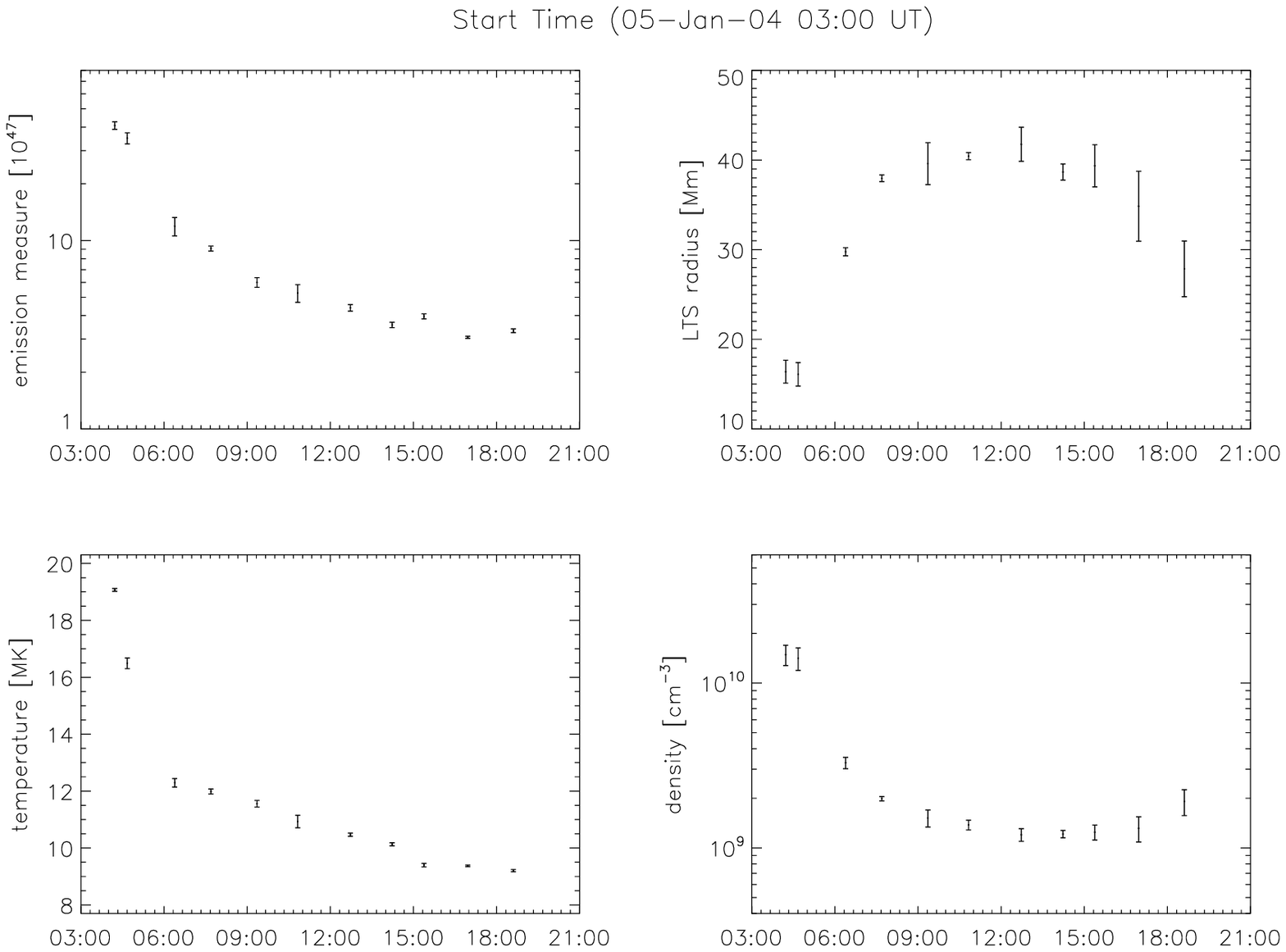,width=12cm}
\end{center}
\caption{Time evolution of observational characteristics obtained for the 2004 January 5 flare. \label{fig4}}
\end{figure}

\section{Observations and data analysis}

The Reuven Ramaty High Energy Solar Spectroscopic Imager ({\it RHESSI}) is a rotating Fourier imager with nine detectors made of pure germanium crystals \citep[]{lin2002}. The detectors record energy and time of arrival for each detected HXR photon. Pairs of grids placed ahead of the detectors and the rotation of the whole satellite (4 s period) cause modulation of HXR radiation coming from actual solar sources. Using several available algorithms we can reconstruct the spatial distribution of the HXR emission from the time-modulated light curves \citep{hurford2002}. 

In our analysis we focused on the very late phase of LDEs when the emission is extremely weak, so some care is needed in the analysis. It is a great difficulty to reconstruct a {\it RHESSI} image when the emitting source is weak. Usually {\it RHESSI} image reconstruction is performed for detectors Nos. 3--6, 8 and 9. Depending on the chosen reconstruction algorithm and weights this set of detectors gives a spatial resolution of the order of 7--9 arcsec \citep[]{aschwanden2002}. However, we were not able to reconstruct any reliable image for the very late phase of an LDE decay. The weak signal was not the cause since taking integration times up to 8 minutes we were able to collect enough counts. 

The problem was resolved in the following way. First, we reconstructed images for single grids and a wide field of view (8~arcsec pixel, image size 256$\times$256 pixels) using the back-projection algorithm \citep[]{hurford2002}. Next, we used these images for determining which grid provides us with reliable images of the source. This step is crucial if we remember  that when a source size is comparable to the resolution of a particular grid then the detector records very weak or no modulation of the signal \citep[]{hurford2002}. In such a case  there is only noise in the fine grids even if the count rates are high. This caused fail in  reconstruction process performed with the use of the standard set of grids. 

In the next step we used only selected coarse grids and reconstructed images with PIXON algorithm \citep[][and references within]{puetter1999}. Since integration times were rather long  (minutes) we used stacked modulations \footnote{http://sprg.ssl.berkeley.edu/$\sim$tohban/nuggets/ ?page=article\&article\_id=39}. The energy ranges chosen for reconstruction were as narrow as possible, i.e. 1~keV hence it was  possible to obtain reliable fits to the spectra derived from the {\it RHESSI} images. The image spectroscopy we performed has an advantage in comparison to "standard" spectroscopy based on the fluxes measured for the whole Sun. Namely, for the rotating Fourier imager the background photons do not influence the modulation profile. Thus, the analysis of very weak fluxes can be done more precisely especially for moments of passage through radiation belts when the background is changing  rapidly with time. 

We did error analysis taking the advantage of slow evolution of LDEs. Namely, we divided each analysed time interval into three intervals. Images were reconstructed for each interval and the  observational characteristics were obtained. Next, we calculated a mean value of each observational characteristic and its standard deviation. We treated the standard deviation as the uncertainty of the obtained characteristics. This method was used for each parameter: size, location, emission measure and temperature.

\section{Results} 
We investigated a set of 9 flares listed in the Table~\ref{tab1}. Fig~\ref{fig1} shows {\it RHESSI} light curve for energy intervals: $6-12$, $12-25$ keV and $7-20$ MeV for the 2004 January 5 flare. The last interval was chosen because it shows clearly passages through radiation belts. The flare HXR emission in the $6-12$~keV band was observed for 15 hours after flare  maximum recorded at 3:45 UT. We marked analysed time intervals with arrows. As mentioned above each interval was divided into three subintervals of the same length.

Reconstructed HXR sources were compared to {\it TRACE} images taken with the use of $171$~\AA \mbox{ } filter.  An example of position of HXR sources ($6-7$~keV) overlaid on {\it TRACE} images is shown in Fig.~\ref{fig2}. The HXR sources were observed above EUV loops and systematically changed their height with velocity of about 2 km/s. We observed no significant footpoint emission in the 2004 January 5 flare and in other events also. 

LDEs are characterized by slow decrease of brightness caused mainly by slow decrease of temperature. Thus, we tried to define an objective method to define the length of the decay phase. For this purpose we fit a Poisson distribution to the observed temperature changes:
\begin{equation}
	T(t)=Ae^{-\frac{t}{\tau}}
\end{equation}
where A is a normalization factor and $\tau$ is a characteristic time. We assumed that the asymptote of this distribution is 2 MK. The characteristic time was estimated for a group of typical, short duration events and we found that it had values of the order of minutes. For the  analysed group of LDEs $\tau$ was never less than 0.5~hour (see Table~\ref{tab2}). 

Obtained observational characteristics of the 2004 January 5 flare are shown in Fig.~\ref{fig4}. Both, temperature and emission measure, decrease with time. For temperature we observed very fast decrease shortly after flare maximum. Next, when temperature reached value 12 MK, this decrease was significantly slower. In the time interval of 6:00 - 18:00 UT temperature changed for 3 MK only. 

The evolution of the radius of the observed source shows two phases. At first the source expanded from $r=16$~arcsec to $r=40$~arcsec at 12:00 UT. After 15:00 UT we observed that source started to diminish to the radius of the order of $28$~arcsec. This observation is very important from methodological point of view. Namely, we chose grids dynamically depending on observed modulation. This result shows that we do not miss information about the actual size when using only coarse grids. 

We observed non-thermal, power-law component in fitted spectra (Fig.~\ref{fig3}). These have to be treated very carefully since slopes are large, close to the value of 10. However, even very steep power-law function is still different from thermal one. We observed non-thermal components in six analysed flares (among 9). It should be mentioned that non-thermal emission was observed in loop-top sources. We did not observe footpoint sources.

\begin{table}[t]
\caption{\footnotesize{\textit{Observational characteristics of analysed LDEs. (1) - Date, (2) - Temperature [MK], (3) - Emission measure $[10^{47}\;\rm{cm}^{-3}]$, (4) - Radius [Mm], (5) - Altitude [Mm], (6) - Density $[10^9\;\rm{cm}^{-3}]$, (7) - Characteristic time of temperature decay [h].}}\label{tab2}}
\begin{center}
\begin{tabular}{cccccccc}
\hline
\noalign{\smallskip}
 & (1) & (2) & (3) & (4) & (5) & (6) & (7)\\
\noalign{\smallskip}
\hline
\noalign{\smallskip}
1 & 25/10/2002 & 14.6-6.7 & 7.9-1.0 & 30-47 & 68-271 & 4.5-0.42 & 13.8\\
2 & 25/8/2003 & 12.3-6.7 & 6.0-0.6 & 5-40 & 54-86 & 17.9-0.5 & 3.2\\
3 & 11/11/2003 & 25.8-7.8 & 1.5-0.3 & 10-48 & 41-103 & 12.8-0.28 & 6.0\\
4 & 05/1/2004 & 26.9-9.0 & 45.0-2.9 & 14-42 & 64-181 & 14.9-1.2 & 31.4\\
5 & 20/1/2005 & 19.0-7.8 & 71.0-3.1 & 17-32 & 13-74 & 18.2-1.8 & 20.0\\
6 & 30/7/2005 & 11.6-7.1 & 3.2-1.9 & 12-21 & 47-60 & 11.0-3.8 & 7.8\\
7 & 22/8/2005 & 12.4-9.9 & 3.8-0.2 & 5-31 & 36-82 & 24.0-1.2 & 6.5\\
8 & 29/11/2005 & 10.1-7.8 & 29.4-0.9 & 12-43 & 29-48 & 15.3-0.16 & 31.4\\
9 & 25/1/2007 & 13.3-9.9 & 13.5-0.2 & 9-18 & 31-73 & 14.0-0.73 & 20.8\\\hline
\end{tabular}
\end{center}
\end{table}

\section{Summary}
We investigated set of 9 LDEs well observed by {\it RHESSI}. In Table~\ref{tab2} we summarized all obtained observational characteristics. For each parameter (except characteristic decay time) we show two values. The first was obtained shortly after flare   maximum and the second was obtained for the last time interval. Such intervals show ranges of values that were observed. The main conclusions of our work are as follow:
\begin{enumerate}
	\item The sensitivity of the {\it RHESSI} detectors allows us to observe the decay phase in HXR emission up to several hours after flare maximum.
	\item Observed loop-top sources are large structures and they usually grow with time. Sometimes we observed decrease of radius after initial growth.
	\item HXR emission sources are located above loops observed in the EUV range. We did not detect any footpoint emission.
	\item Obtained values give strong constraints for models of solar flares since high temperatures demand continuous energy release. 
	\item The observed non-thermal component is another proof for continuing energy release. However it must be stressed that observed power-law components have very steep spectra with indices close to the value of 10.
\end{enumerate}

\section*{Acknowledgements} 
The {\it RHESSI} satellite is NASA Small Explorer (SMEX) mission. We acknowledge many useful and inspiring discussions of Professor Micha\l \mbox{ }Tomczak. We also thank Barbara Cader-Sroka for editorial remarks. This investigation has been supported by a Polish Ministry of Science and High Education, grant No. N203 1937 33.
\bibliographystyle{ceab}


\end{document}